\documentclass[12pt]{iopart}
\usepackage{graphicx}
\usepackage{dcolumn}
\usepackage{bm}
\usepackage{color}

\begin{document}

\title[]{Effects of higher levels of qubits on control of qubit protected by a Josephson quantum filter}

\author{Shumpei Masuda$^{1,2}$ and Kazuki Koshino$^{1}$}
\address{$^{1}$ College of Liberal Arts and Sciences, Tokyo Medical and Dental University, Ichikawa, 272-0827, Japan}
\address{$^{2}$ National Institute of Advanced Industrial Science and Technology, Tsukuba, Ibaraki 305-8565, Japan}
\ead{shumpei.masuda@aist.go.jp}

\vspace{10pt}

\begin{abstract}
A Josephson quantum filter (JQF)  protects a data qubit (DQ) from the radiative decay into transmission lines in superconducting quantum computing architectures.
A transmon, which is a weakly nonlinear harmonic oscillator rather than a pure two-level system, can play a role of a JQF or a DQ. 
However, in the previous study, a JQF and a DQ were modeled as two-level systems neglecting the effects of higher levels.
We theoretically examine the effects of the higher levels of the JQF and the DQ on the control of the DQ.
It is shown that the higher levels of the DQ cause the shift of the resonance frequency and the decrease of the maximum population of the first excited state of the DQ in the controls with a continuous wave (cw) field and a pulsed field, while the higher levels of the JQF do not.
Moreover, we present optimal parameters of the pulsed field, which maximize the control efficiency. 
\end{abstract}

%
%
%
%
%

\section{Introdunction}
In waveguide quantum electrodynamics (QED) systems, an atom is coupled strongly to a one-dimensional (1D) optical field typically provided by a waveguide or a transmission line (TL), so that spontaneous emission from an atom is mostly forwarded to this one-dimensional field.
Such systems are indispensable for realization of distributed quantum computation, in which photonic qubits quantum-mechanically connect distant matter qubits. In contrast with the natural atom-atom interaction, which becomes weaker rapidly as their mutual distance increases, the atom-atom interaction in waveguide QED systems is long-ranged owing to the one-dimensionality of the field.

A waveguide QED system was first realized with a cavity QED system (atom-cavity coupled system) in the bad-cavity regime exploiting the Purcell effect \cite{Turchette1995}. 
Waveguide QED systems can be realized also in superconducting circuits, which is a  promising platform for quantum information processings \cite{Nakamura1999,Blais2004,Wallraff2004,Astafiev2007,Majer2007,Sillanpaa2007,Astafiev2010,Devoret2013,Kelly2015,Ofek2016}, by coupling a superconducting artificial atom directly to a microwave TL \cite{Astafiev2010,Shen2005,Hoi2011}.
This enabled us to implement a waveguide QED setup involving several atoms coupled to a common waveguide \cite{Loo2013}. In such setups, distant atoms can interact with each other via virtual photons propagating in the waveguide. 
The coupling between a superconducting artificial atom with a 1D waveguide has been achieved even in the ultrastrong coupling regime \cite{FD2017}.
Quantum computation schemes \cite{Zheng2013,Paulisch2016}, two-photon nonlinearlities and photon correlation function \cite{Fang2015} were studied in waveguide QED systems.

Unwanted radiative decay of qubits degrades quantum compuation.
Various methods to decrease or design qubit decay  in circuit QED systems have been studied using, $e.g.$, effect of boundary condition \cite{Koshino2012},  mirror \cite{Hoi2015}, other multiple qubits \cite{Loo2013,Chang2012,Lalumiere2013,Mirhosseini2019} including superconducting metamaterials \cite{Mirhosseini2018}.

It was shown that a qubit attached to a TL with suitable parameters can work as a filter, which prohibits a data qubit (DQ) from radiative decay to the TL \cite{Koshino2020,Kono2020}.
The protecting qubit is called a Josephson quantum filter (JQF).
A transmon can work as a JQF or a DQ in superconducting quantum computing architectures.
A transmon is a weakly nonlinear harmonic oscillator rather than a pure two-level system \cite{Koch2007}.
However, in the previous study \cite{Koshino2020}, both the DQ and the JQF were modeled as pure two-level systems neglecting higher levels.

In this paper, we consider controls of the DQ with a cw field and a pulsed field, which are routinely performed for calibrations of experimental apparatuses, parameter determinations, and quantum information processing.
We examine the effects of the higher levels of the qubits and show the shift of the resonance frequency and the change in the maximum fidelity of the controls induced by them.
Furthermore, we show optimal parameters for controls with a pulsed field.

The rest of this paper is organized as follows.
In Sec.~\ref{Model}, we introduce a model for the system.
In Sec.~\ref{Effect of a higher level in cw drive}, we derive formulae of the resonance frequency and the maximum population of the first excited state of the DQ under a cw field.
We numerically study the controls of the DQ with a cw field and a pulsed field in Sec.~\ref{Numerical results}. The results are compared with the theoretical prediction. We present an optimal pulse length for the control with a pulsed field.
Section~\ref{Summary} provides a summary.

\section{Model}
\label{Model}

\begin{figure}
\begin{center}
\includegraphics[width=9cm]{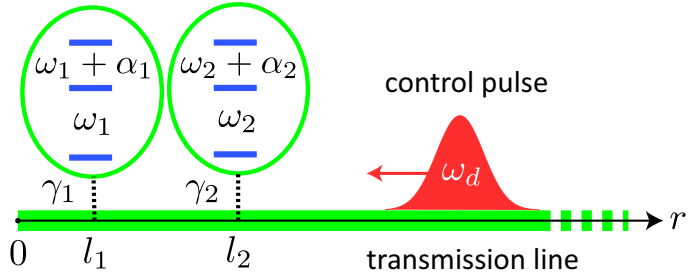}
\end{center}
\caption{
Schematic of the setup.
The DQ and the JQF are coupled to a
semi-infinite TL, through which control pulses for the DQ is applied.
}
\label{schematic_3_13_20}
\end{figure}

Our system is composed of two qubits, the DQ (qubit 1) and the JQF (qubit 2),  attached to a semi-infinite TL, which extends in the $r>0$ region.
The schematic of the setup is illustrated in Fig.~\ref{schematic_3_13_20}.
The position, angular frequency, anharmonicity parameter and coupling strength to the TL of qubit $m(=1,2)$ are denoted by $l_{m}$, $\omega_m$, $\alpha_m$ and $\gamma_m$, respectively. 
When $l_1<l_2$ and $\gamma_1\ll\gamma_2$, qubit 2 can work as a JQF, which prohibits the radiative decay of qubit 1  \cite{Koshino2020}.
In this study, we assume that the resonance frequencies of the qubits are identical and that the positions of the qubits are optimal, that is, $l_1=0$ and $l_2/\lambda_q=0.5$,
where $\lambda_q$ is the resonance wavelength of the qubits.

Adopting the units in which $\hbar=v=1$, where $v$ is the microwave velocity in the TL, the Hamiltonian of the system is represented as
\begin{eqnarray}
H &=& \sum_m \Big{(} \omega_m c^\dagger_m c_m +
\frac{\alpha_m}{2} c^\dagger_m c^\dagger_m c_m c_m \Big{)}
\nonumber\\
&&+\int_0^\infty dk \Big{[} k b_k^\dagger b_k  
+ \sum_m g_{mk} (c_m^\dagger b_k + b_k^\dagger c_m) \Big{]},
\label{H_12_3_19}
\end{eqnarray}
where $c_m$ is the annihilation operator of qubit $m$, and
$b_{k(>0)}$ is the annihilation operator of the eigenmode of the TL with the wave number $k$ and the mode function, $f_k=\sqrt{2/\pi} \cos kr$, normalized as $\int_0^\infty dr f_{k'}(r) f_k(r)=\delta(k-k')$.
The coupling constant between qubit $m$ and the TL is given by
\begin{eqnarray}
g_{mk} = \sqrt{\frac{\gamma_m}{2}} f_k(l_m) = \sqrt{\frac{\gamma_m}{\pi}} \cos(kl_m).
\label{g_10_10_19}
\end{eqnarray}

\subsection{Equation of motion}
The Heisenberg equation for $b_k$ leads to
\begin{eqnarray}
\frac{d}{dt}b_k = -ikb_k
-i\sum_m g_{mk} c_m,
\end{eqnarray}
which is formally solved as
\begin{eqnarray}
b_k(t) = b_k(0) e^{-ikt} - i \sum_m g_{mk} \int_0^t dt' c_m(t') e^{ik(t'-t)}.
\label{bk_3_13_20}
\end{eqnarray}
We formally extend the lower limit of $k$ to $-\infty$
in order to introduce the real-space representation of the field operator defined by 
\begin{eqnarray}
\tilde{b}_r = \frac{1}{\sqrt{2\pi}} \int_{-\infty}^{\infty} dk\ e^{ikr} b_k.
\label{br_10_10_19}
\end{eqnarray}
Here, $r$ runs over $-\infty < r < \infty$.
The negative and positive regions represent the incoming and outgoing fields, respectively.
The introduction of the real-space representation has been validated in Ref.~\cite{Banacloche2013}.
Using Eqs.~(\ref{bk_3_13_20}) and (\ref{br_10_10_19}), we obtain 
\begin{eqnarray}
\tilde{b}_r(t) &=& \tilde{b}_{r-t}(0) -
i\sum_m \sqrt{\frac{\gamma_m}{2}} \Big{[}
\Theta_{r\in (-l_m,t-l_m)} c_m (t-r-l_m) \nonumber\\
&&+ \Theta_{r\in (l_m,t+l_m)} c_m (t-r+l_m) \Big{]}, 
\label{Heisenberg_12_3_19}
\end{eqnarray}
where $\Theta_{r\in (a,b)}=\theta(r-a)\theta(b-r)$.
Using Eq.~(\ref{Heisenberg_12_3_19}), we can obtain 
\begin{eqnarray}
\tilde{b}_{l_m}(t) + \tilde{b}_{-l_m}(t) &=& \tilde{b}_{l_m-t}(0) + \tilde{b}_{-l_m-t}(0)-i \sum_n \sqrt{\frac{\gamma_n}{2}} \big{[} c_n(t-l_m-l_n) \nonumber\\
&& + c_n(t- |l_m-l_n| )\big{]}.
\label{blm_12_3_19}
\end{eqnarray}

On the other hand, the Heisenberg equation for a system operator $O$ composed of qubit operators is written as
\begin{eqnarray}
\frac{d}{dt}O &=& i[H_s,O] + 
i\sum_m \sqrt{\frac{\gamma_m}{2}} \Big( [c_m^\dagger,O] \Big\{ \tilde{b}_{l_m}(t) + \tilde{b}_{-l_m}(t) \Big\}
\nonumber\\
&&+ \Big\{ \tilde{b}_{l_m}^\dagger(t) + \tilde{b}_{-l_m}^\dagger(t) \Big\} [c_m,O] \Big{)},
\label{dsdt_12_3_19}
\end{eqnarray}
where $H_s = \sum_m (\omega_m c_m^\dagger c_m + \frac{\alpha_m}{2} c_m^\dagger c_m^\dagger c_m c_m)$.
Substitution of Eq.~(\ref{blm_12_3_19}) into Eq.~(\ref{dsdt_12_3_19}) leads to
\begin{eqnarray}
\frac{d}{dt} O &=&  i[H_s,O] + i \sum_m \Big\{ [ c_m^\dagger, O] N_m(t) + N_m^\dagger(t)[ c_m, O ] \Big\} \nonumber\\
&& + \sum_{m,n} \frac{\sqrt{\gamma_m\gamma_n}}{2} [c_m^\dagger,O]
\Big\{ c_n(t-l_m-l_n) + c_n(t- |l_m-l_n|) \Big\} 
\nonumber \\
&&
- \sum_{m,n} \frac{\sqrt{\gamma_m\gamma_n}}{2} \Big\{
c_n^\dagger(t-l_m-l_n) + c_n^\dagger(t-| l_m-l_n |) \Big\} [c_m,O],
\label{dodt_12_3_19}
\end{eqnarray}
where $N_m(t)$ is the noise operator defined by
\begin{eqnarray}
N_m(t) = \frac{\gamma_m}{2} \Big{[} \tilde{b}_{l_m-t}(0) + \tilde{b}_{-l_m-t}(0) \Big{]}.
\label{N_3_16_20}
\end{eqnarray}
By replacing $c_m(t-\Delta t)$ with $e^{i\omega_m \Delta t}c_m$ (free evolution approximation~\cite{Koshino2020}), the equation of motion is rewritten as
\begin{eqnarray}
\frac{d}{dt}O &=& i [H_s, O] + i\sum_m \Big\{ [c_m^\dagger,O] N_m(t) + N_m^\dagger(t) [c_m,O] \Big\}
\nonumber\\
&& + \sum_{m,n} \Big{(} \xi_{mn} [c_m^\dagger,O] c_n - \xi_{mn}^\ast c_n^\dagger [c_m,O] \Big{)},
\label{dodt2_12_6_19}
\end{eqnarray}
where
\begin{eqnarray}
\xi_{mn} = \frac{\sqrt{\gamma_{m} \gamma_{n}}}{2} \Big{(} e^{i\omega_q (l_m+l_n)} 
+ e^{i\omega_q |l_m-l_n|} \Big{)},
\label{xi_10_15_19}
\end{eqnarray}
where $\omega_q=\omega_1=\omega_2$.

\subsection{Dynamics under control field}
We assume that the qubits are in the ground state at the initial time, and that a classical control field $E_{\rm in}(t)$ is applied for $t>0$. 
The spatial waveform of the control field at $t=0$ is represented as $E_{\rm in}(-r)$.
The initial state vector is written as
\begin{eqnarray}
|\phi(0)\rangle = N \exp \Big{(} \int_{-\infty}^0 dr\ E_{\rm in}(-r) \tilde{b}_r^\dagger \Big{)} |v\rangle,
\label{phi0_12_4_19}
\end{eqnarray}
where $N=\exp(-\int dr |E_{\rm in}(-r)|^2/2)$ is a normalization factor, and $|v\rangle$ is the overall ground state, the product state of the ground states of two qubits and the vacuum states of the waveguide modes.
The initial state is an eigenstate of the noise operator in Eq.~(\ref{N_3_16_20}) because it is in a coherent state.
We can calculate the time evolution of the density matrix of the system under the control field based on this equation of motion in Eq.~(\ref{dodt2_12_6_19}).
Note that $N_m$ in Eq.~(\ref{dodt2_12_6_19}) can be replaced by 
\begin{eqnarray}
\langle N_1(t) \rangle = \sqrt{2\gamma_1} \cos(\omega_1 l_1) E_{\rm in}(t)\nonumber\\
\langle N_2(t) \rangle = \sqrt{2\gamma_2} \cos(\omega_2 l_2) E_{\rm in}(t)
\end{eqnarray}
where we used the notation of $\langle A(t) \rangle = \langle \phi(0)| A(t) | \phi(0)\rangle$.
The control field is represented as 
\begin{eqnarray}
E_{\rm in}(t) = 2 E_d(t) \cos(\omega_d t)
\end{eqnarray}
where the frequency and the envelope of the control field are $\omega_d/2\pi$ and $E_d$, respectively.

\section{Effects of a higher level in cw drive}
\label{Effect of a higher level in cw drive}
We consider a cw drive of the DQ protected by the JQF.
As shown in the following section, we observe the shift of the resonance frequency and the decrease of maximum population of the first excited state in Rabi oscillations.
We attribute these to the second excited state of the DQ, and
derive analytic formulae of the resonance frequency and the maximum population with the use of an effective Hamiltonian, which consists of a transmom under a control field.

The effective time-dependent Hamiltonian describing the DQ is given by
\begin{eqnarray}
H(t) =  \omega c^\dagger c +
\frac{\alpha}{2} c^\dagger c^\dagger c c +
2\Omega\cos(\omega_d t) (c^\dagger + c),
\end{eqnarray}
where 
$\omega = \omega_q$, $\alpha=\alpha_1 < 0$ and $c=c_1$.
Here, $\Omega$ is the Rabi frequency, which is related to the control field by 
 $\Omega=\sqrt{2\gamma_1}E_d$.
We assume that the effects of the JQF is negligible, when investigating the dynamics under a cw drive field.
Now, we consider a subsystem spanned by three levels $|0\rangle,|1\rangle$ and $|2\rangle$.
The Hamiltonian is represented as 
\begin{eqnarray}
H =    \left( \begin{array}{ccc}
0  & 2\Omega \cos(\omega_d t) & 0    \\
2\Omega \cos(\omega_d t) & \omega  & 2\sqrt{2}\Omega \cos(\omega_d t) \\
0 & 2\sqrt{2}\Omega \cos(\omega_d t) & 2\omega + \alpha \end{array}
 \right).
\end{eqnarray}
We move to a rotating frame with angular frequency of $\omega_d$ and use the rotating wave approximation to rewrite the Hamiltonian as 
\begin{eqnarray}
H =   \left( \begin{array}{ccc}
0  & \Omega  & 0    \\
\Omega  & \omega-\omega_d  & \sqrt{2}\Omega  \\
0 & \sqrt{2}\Omega  & 2 (\omega-\omega_d)+\alpha \end{array}
 \right).
\label{H_ori_3_16_20}
\end{eqnarray}
We consider a subspace expanded by $|1\rangle,|2\rangle$ in which the Hamiltonian is represented as
\begin{eqnarray}
H_2 &=&   \left( \begin{array}{cc}
 \omega-\omega_d  & \sqrt{2}\Omega  \\
 \sqrt{2}\Omega  & 2 (\omega-\omega_d)+\alpha \end{array}
 \right)\nonumber\\
 &=& \frac{3(\omega-\omega_d) + \alpha}{2} I_2
 + \left( \begin{array}{cc}
a & b  \\
b  & -a \end{array}
 \right),
\end{eqnarray}
where $I_2$ is the identity operator and 
\begin{eqnarray}
a &=& \frac{-\alpha-\omega+\omega_d}{2}, \nonumber\\
b &=& \sqrt{2}\Omega.
\end{eqnarray}
The eigenenergies are represented as
\begin{eqnarray}
E_\pm = \frac{3(\omega-\omega_d) + \alpha}{2} \pm \sqrt{a^2 + b^2},
\label{Ep_3_16_20}
\end{eqnarray}
and the corresponding eigenstates are written as
\begin{eqnarray}
|\pm \rangle = \cos\theta_{\pm} |1\rangle + \sin\theta_{\pm} |2\rangle,
\end{eqnarray}
where 
\begin{eqnarray}
\tan\theta_\pm = \frac{a\pm \sqrt{a^2 + b^2}}{b}.
\end{eqnarray}
Alternatively, the eigenstates are represented as 
\begin{eqnarray}
|\pm \rangle = - \frac{b}{s} |1\rangle
+ \frac{a \mp \sqrt{a^2 + b^2}}{s} |2\rangle
\label{p1_3_16_20}
\end{eqnarray}
with
\begin{eqnarray}
s = \Big\{ 2 \sqrt{a^2 + b^2} (\sqrt{a^2 + b^2} - a) \Big\}^{1/2}.
\end{eqnarray}

We rewrite the Hamiltonian in Eq.~(\ref{H_ori_3_16_20}) with the basis set $\{ |0\rangle,|+\rangle,|-\rangle \}$.
In the matrix representation, the Hamiltonian is represented as
\begin{eqnarray}
H =   \left( \begin{array}{ccc}
0  & \Omega\cos\theta_+  & \Omega\cos\theta_-    \\
\Omega\cos\theta_+  & E_+  & 0  \\
\Omega\cos\theta_- & 0  & E_- \end{array}
 \right).
\end{eqnarray}
Note that $|+\rangle\simeq |1\rangle$ when $\Omega/|\alpha|$ is sufficiently small. 
We also emphasize that $\cos\theta_+ \simeq 1 \gg \cos\theta_-$. 
Therefore, $|-\rangle$ can be neglected, and
a Rabi oscillation will be observed between 
$|0\rangle$ and $|+\rangle$ when parameters are chosen so that $E_+=0$. 
We regard $E_+=0$ as the resonance condition.
For example, $\omega_d$ can be tuned to satisfy the resonance condition.
The deviation of $\omega_d$ from $\omega$ is the origin of the shift of the resonance frequency.

Now we derive an analytic form of the shift of the resonance frequency.
When $\Omega/|\alpha|\ll 1$, we have $a\gg b$. 
Then, we obtain from Eq.~(\ref{Ep_3_16_20})
\begin{eqnarray}
E_+ &\simeq&  \omega - \omega_d 
+ \frac{2\Omega^2}{- \alpha - \omega + \omega_d},
\label{Ep_7_21_20}
\end{eqnarray}
where we used $\sqrt{a^2 + b^2} \simeq a + b^2/2a$. 
Thus, we obtain
\begin{eqnarray}
\omega_d \simeq \omega + \frac{2\Omega^2}{ - \alpha - \omega + \omega_d},
\end{eqnarray}
when $E_+=0$.
Assuming $\omega_d\simeq \omega$, we can derive a simple resonance condition
\begin{eqnarray}
\omega_d  - \omega  \simeq - \frac{2\Omega^2}{\alpha}.
\label{shift_3_17_20}
\end{eqnarray}
This represents the shift of the resonance frequency as a function of $\Omega$ and $\alpha$.

Equation~(\ref{p1_3_16_20}) shows that the maximum population of $|1\rangle$ during the Rabi oscillation between 
$|0\rangle$ and $|+\rangle$ is given by 
\begin{eqnarray}
p_{1,{\rm cw}}^{\rm max} = \frac{b^2}{s^2}.
\label{p1_3_17_20}
\end{eqnarray}
By using $\Omega/|\alpha|\ll 1$ and $a\simeq -\alpha/2$, an approximate form of the maximum population can be derived as 
\begin{eqnarray}
p_{1,{\rm cw}}^{\rm max} \simeq 1 - \frac{2\Omega^2}{\alpha^2}.
\label{p1_3_17_20_ver2}
\end{eqnarray}
The same formula can be obtained using Schrieffer-Wolff transformation (see Appendix \ref{Analysis with Schrieffer-Wolff transformation}).

\section{Numerical results}
\label{Numerical results}
In this section, we numerically examine controls of the DQ with a cw field and with a gaussian pulse
focusing on the shift of the resonance frequency and the maximum value of the population, $p_1$, of the first excited state of the DQ.
The numerical results are compared with the theoretical prediction in Sec.~\ref{Effect of a higher level in cw drive} for the control with a cw field.
An optimal pulse length is presented for the control with a gaussian pulse.
It is also shown that the maximum value of $p_1$ for the control with a gaussian pulse is higher than the one for the control with a cw field.

\subsection{cw drive}
\begin{figure}
\begin{center}
\includegraphics[width=8cm]{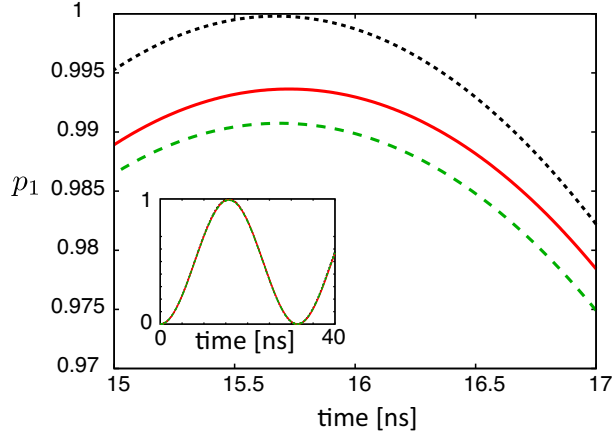}
\end{center}
\caption{
Time dependence of the population of the first excited state of the DQ, $p_1$, under the cw control field.
The used parameter set is $\alpha/2\pi=-300$~MHz, $\omega_{1,2}/2\pi=5$~GHz, $\gamma_1/2\pi=2$ kHz, $\gamma_2/2\pi=100$ MHz and $\sqrt{2\gamma_1}E_d/2\pi=16$~MHz.
Higher levels are taken into account for the data represented by the red solid and the green dashed curves, while the black dotted curve corresponds to the system where the qubits are modeled as two-level systems. The frequency of the control field is $\omega_d/2\pi=5$~GHz for the green dashed and the black dotted curves, while $\omega_d/2\pi=5.0017$~GHz for the red solid curve, which is optimized to give the highest value of $p_1$.
Inset: the same plot for a wider time range. Three lines are mostly overlapping.
 }
\label{pop_com_3_16_20}
\end{figure}

We simulate the dynamics of the system under a cw control field and calculate the population of the first excited state of the DQ defined by $p_1=\langle {\Pi}_1 \otimes {I} \rangle$,
where ${\Pi}_1$ is the projection operator to the first excited state of the DQ and 
$I$ denotes the identity operator for the JQF.
Figure~\ref{pop_com_3_16_20} shows the time dependence of $p_1$. 
As a reference, we also calculate the dynamics of the system, where both of the DQ and the JQF are modeled as two-level systems, under the cw field with $\omega_d=\omega_q$ (dotted line in Fig.~\ref{pop_com_3_16_20}).
The maximum value of $p_1$ is slightly less than unity due to the effects of the JQF \cite{Koshino2020}.
On the other hand, the maximum value of $p_1$ for the system in which higher levels are taken into account
is further lowered even if the frequency of the control field is optimized (solid line in Fig.~\ref{pop_com_3_16_20}). 

\begin{figure}
\begin{center}
\includegraphics[width=8cm]{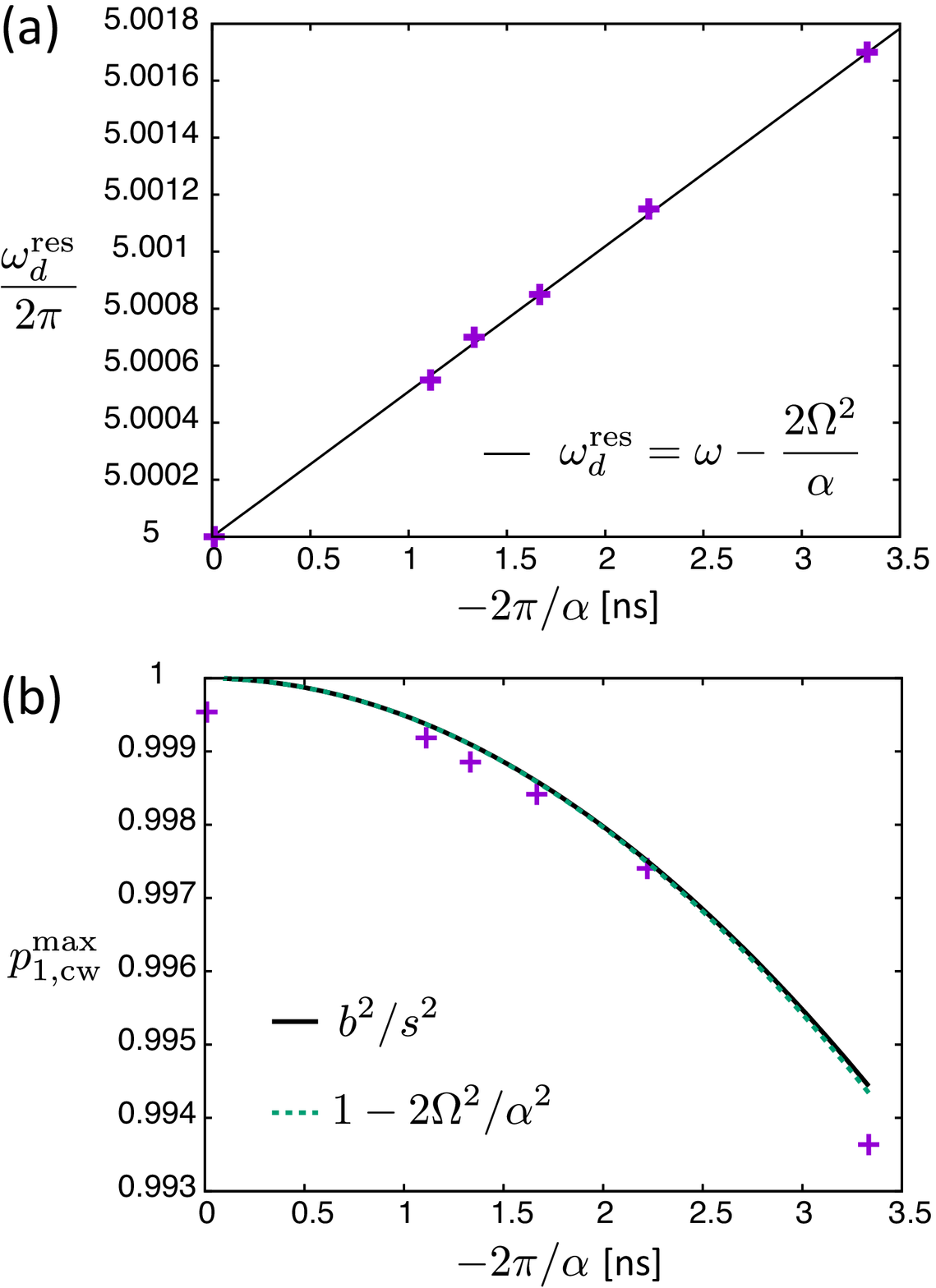}
\end{center}
\caption{
(a) Resonance frequency, $\omega_d^{\rm res}/2\pi$,  for cw drive for various values of $\alpha$.
The solid line corresponds to the theoretical prediction in Eq.~(\ref{shift_3_17_20}).
(b) Maximum value of $p_1$ for various values of $\alpha$.
The solid and the dotted curves, which are almost overlapping, represent theoretical prediction in Eqs.~(\ref{p1_3_17_20}) and (\ref{p1_3_17_20_ver2}), respectively.
The other parameters used are the same as those in Fig.~\ref{pop_com_3_16_20}.
}
\label{fd_cw_3_5_20}
\end{figure}

The optimized frequency of the control field, $\omega_d^{\rm res}/2\pi$, which maximizes $p_1$, deviates from the resonance frequency of a bare qubit.
Figure~\ref{fd_cw_3_5_20}(a) shows $\omega_d^{\rm res}/2\pi$ as a function of $\alpha$.
The other parameters used are the same as those in Fig.~\ref{pop_com_3_16_20}.
It is observed that $\omega_d^{\rm res}/2\pi$ increases linearly with respect to $-1/\alpha$ and is consistent with the analytic expression in Eq.~(\ref{shift_3_17_20}).

Figure~\ref{fd_cw_3_5_20}(b) shows the maximum value of $p_1$ denoted by $p_{1,{\rm cw}}^{\rm max}$ as a function of $1/\alpha$.
It is seen that $p_{1,{\rm cw}}^{\rm max}$ decreases when the anharmonicity parameter $-\alpha$ decreases because the higher excited states become more populated. 
The numerical result agrees with the theoretical prediction in Eqs.~(\ref{p1_3_17_20}) and (\ref{p1_3_17_20_ver2}) although $p_{1,{\rm cw}}^{\rm max}$ is slightly lower than the theoretical result.
We attribute the difference between the numerical and the theoretical result to the finite coupling between $|0\rangle$
and $|-\rangle$.
The JQF and the levels of the DQ higher than its second excited state also contribute to the difference 
because the difference becomes smaller when they are omitted.

\begin{figure}
\begin{center}
\includegraphics[width=7cm]{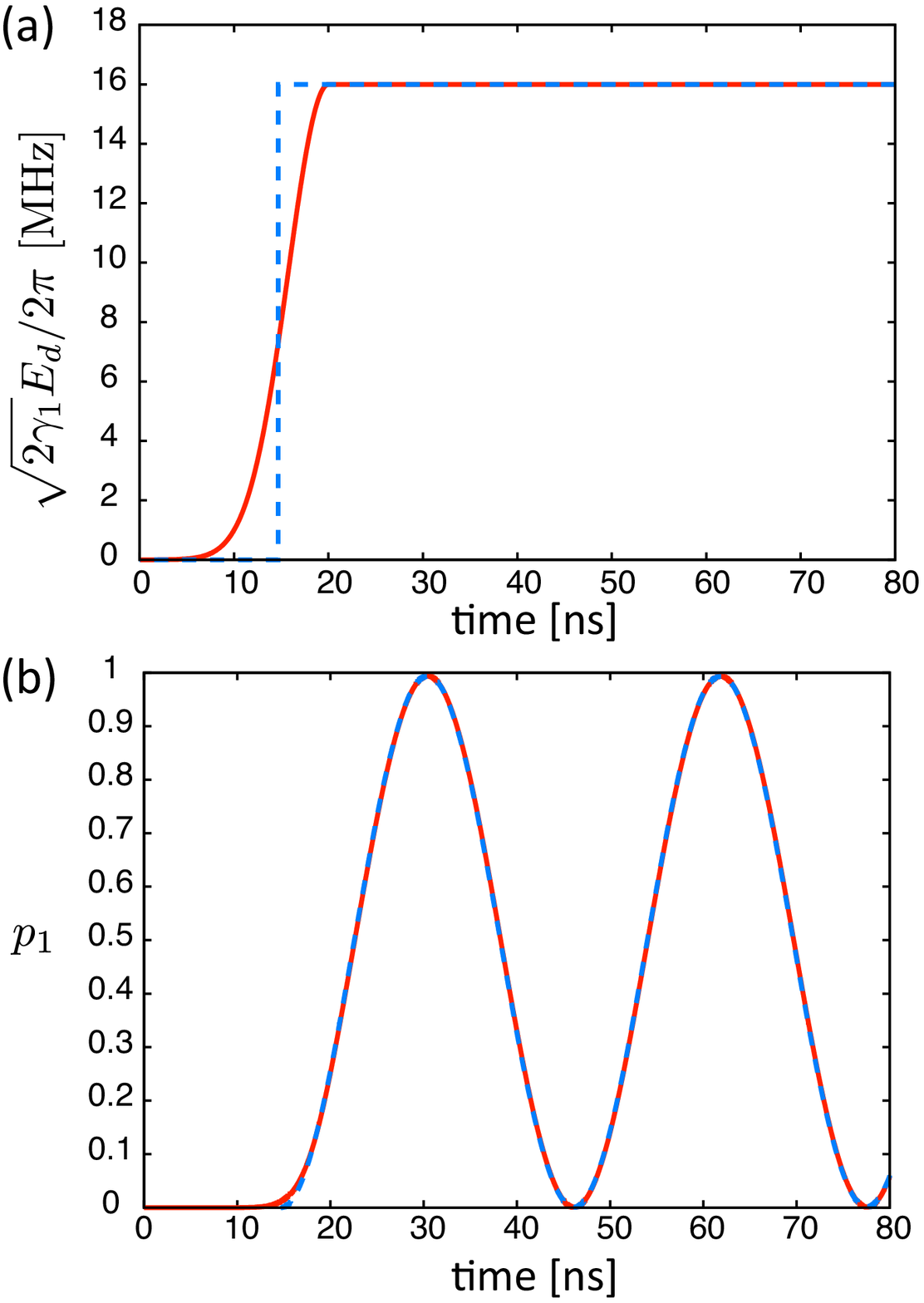}
\end{center}
\caption{
(a) Time profiles of the drive field: Eq.~(\ref{Eamp_5_2_20}) (red solid curve) and Eq.~(\ref{Ed_6_27_20}) (blue dashed curve).
(b) Time dependences of $p_1$ under the drive with $E_d$ in Eq.~(\ref{Eamp_5_2_20}) (red solid curve) and Eq.~(\ref{Ed_6_27_20}) (blue dashed curve).
We set $\sqrt{2\gamma_1}E_{\rm amp}/2\pi=16$~MHz, $\omega_d/2\pi=5.0017$~GHz, $\sigma=10$~ns, $t_0=20$~ns and $t_0'=14.7$~ns.
The other parameters used are the same as those in Fig.~\ref{pop_com_3_16_20}.
}
\label{pop_Ed_cw_4_5_20}
\end{figure}

Similar decrease of the maximum value of $p_1$ occurs even if the drive amplitude, $E_d$, is gradually increased.
We consider the case in which $E_d$ is increased with a gaussian form and becomes constant.
$E_d$  is  represented as 
\begin{eqnarray}
E_d(t) = \left\{ 
\begin{array}{cc}
E_{\rm amp} \exp \Big{(} - 4\ln2\frac{(t-t_0)^2}{{\rm \sigma}^2}\Big{)}\ & {\rm for} \ t<t_0, \\
E_{\rm amp} \ & {\rm for} \ t\ge t_0.
\end{array}
\right.
\label{Eamp_5_2_20}
\end{eqnarray}
The time dependence of $E_d$ is shown in Fig.~\ref{pop_Ed_cw_4_5_20}(a).
Figure~\ref{pop_Ed_cw_4_5_20}(b) shows the time dependences of $p_1$ under the drive with $E_d$ in Eq.~(\ref{Eamp_5_2_20}) and the drive with 
\begin{eqnarray}
E_d(t) = E_{\rm amp} \theta(t-t_0'),
\label{Ed_6_27_20}
\end{eqnarray}
where $\theta$ is the Heaviside step function.
Here, $t_0'$ is set so that the pulse areas of the both controls are the same.
The maximum values of $p_1$  in the both controls are approximately 0.994.

\subsection{pulsed drive}
\begin{figure}
\begin{center}
\includegraphics[width=8cm]{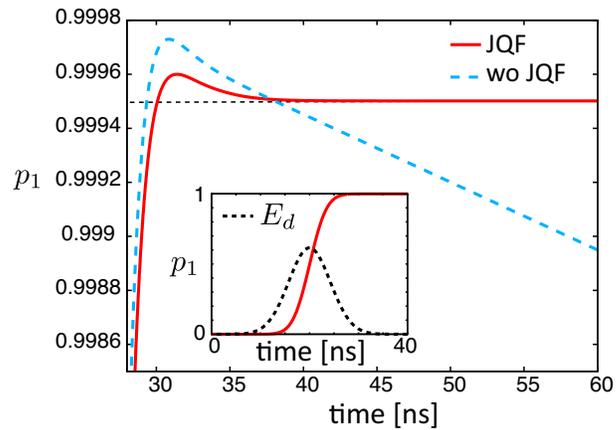}
\end{center}
\caption{
Time dependence of $p_1$ under the drive with a gaussian pulse with $\sigma=10$~ns and $t_0=20$~ns. 
The red solid curve is for the full model, which takes into account the higher levels and the JQF.
The blue dashed curve is for the system without JQF.
The other parameters used are the same as those in Fig.~\ref{pop_com_3_16_20}.
The four lowest levels of DQ and the two lowest levels of the JQF are taken into account.
The horizontal dotted line represents $p_{1,p}^{\rm opt}$.
The inset shows $p_1$ and the pulse envelope in an arbitrary unit for $0\le t\le 40$~ns.
}
\label{dyn_gauss_3_17_20}
\end{figure}

We consider a $\pi$ pulse control aiming at a bit flip of the DQ from the ground state.
In this study, we consider a gaussian pulse represented as
\begin{eqnarray}
E_d(t) =  E_{\rm amp} \exp \Big{(} - 4\ln2\frac{(t-t_0)^2}{{\rm \sigma}^2}\Big{)},
\label{Ed_gauss_5_4_20}
\end{eqnarray}
where $t_0$, $\sigma$, $E_{\rm amp}$ are the pulse center, full width at half maximum and the height of the pulse, respectively. 
Figure~\ref{dyn_gauss_3_17_20} shows the time dependence of $p_1$ during the control.
The frequency and the amplitude of the drive field are optimized for $\sigma$ of 10 ns to maximize $p_1$.
$p_1$ is increased up to 0.9995 in spite of the existence of the higher levels, and it becomes stationary after the control because the JQF prohibits unwanted radiative decay of the DQ (solid line in Fig.~\ref{dyn_gauss_3_17_20}).
This insensitivity of the control efficiency to the higher levels is attributed to the narrow distribution of the pulse field in the frequency space.
The full width at half maximum of the pulse field in the frequency space is approximately 88~MHz, and it is smaller than the absolute value of the anharmonicity parameter.
The comparison with the result for the system without JQF in which $p_1$ decreases exponentially with time due to the radiative decay to the TL after the pulse injection, highlights the protection of the DQ by the JQF (dotted line in Fig.~\ref{dyn_gauss_3_17_20}).

\begin{figure}
\begin{center}
\includegraphics[width=8cm]{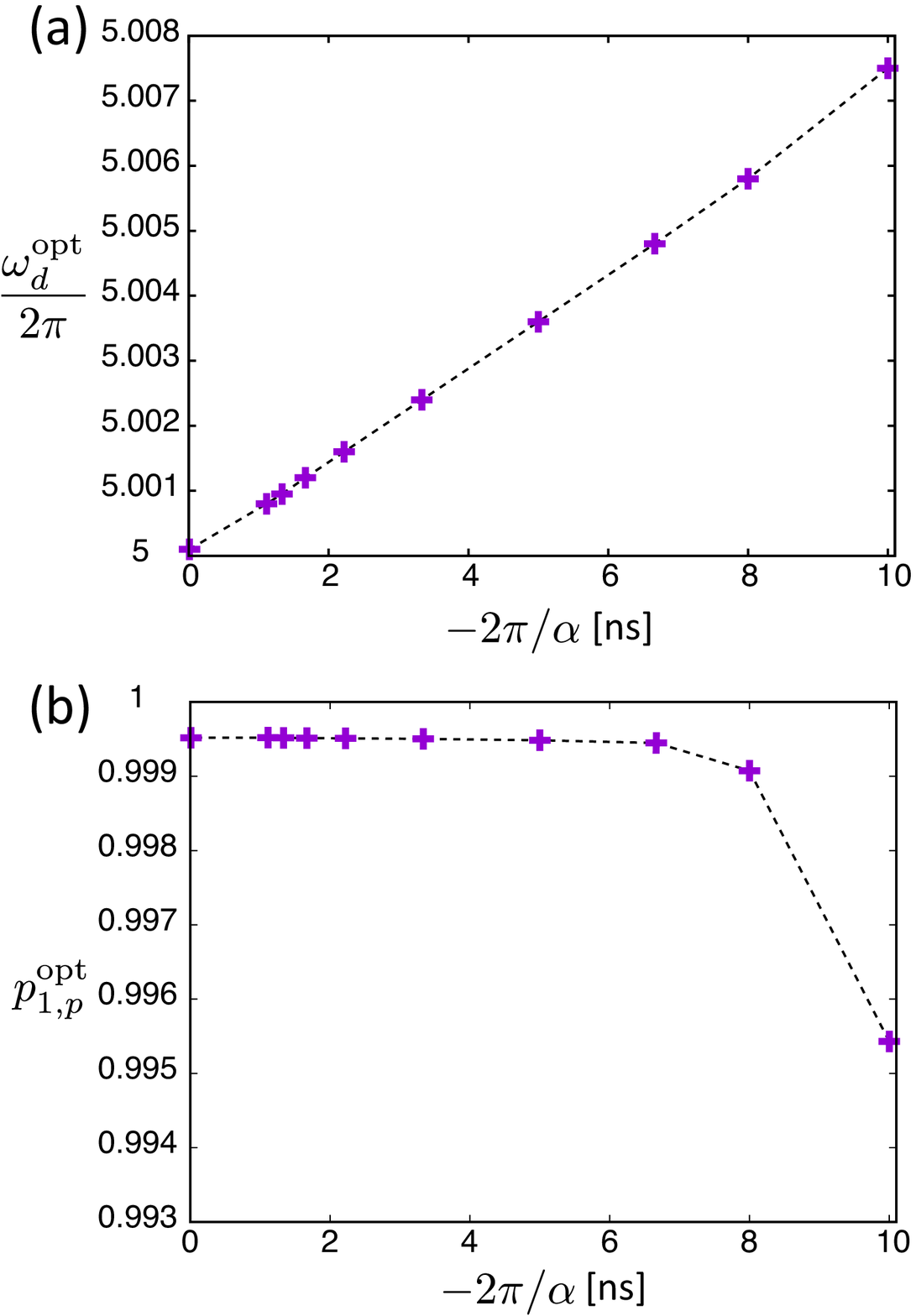}
\end{center}
\caption{
(a) Optimal drive frequency, $\omega_d^{\rm opt}/2\pi$,  for a pulsed drive with $\sigma=10$~ns for various values of $\alpha$.
The dotted line is  guide to the eye.
(b) $p_{1,p}^{\rm opt}$ for various values of $\alpha$.
The other parameters used are the same as those in Fig.~\ref{pop_com_3_16_20}.
}
\label{res_fr_pmax_pulse_3_17_20}
\end{figure}

The optimal drive frequency, $\omega_d^{\rm opt}/2\pi$, which maximizes $p_1$, is shown as a function of $\alpha$ in Fig.~\ref{res_fr_pmax_pulse_3_17_20}(a).
The shift of the optimal drive frequency increases almost linearly with respect to $-1/\alpha$ similar to the case with the cw drive.
The maximum value of $p_1$ after the $\pi$ pulse control, $p_{1,p}^{\rm opt}$, in Fig.~\ref{res_fr_pmax_pulse_3_17_20}(b) is insensitive to $-1/\alpha$ for $-2\pi/\alpha < 5$~ns and is high compared to the case of the cw drive in Fig.~\ref{fd_cw_3_5_20}. 
However, $p_{1,p}^{\rm opt}$ decreases as $-1/\alpha$ increases further because higher levels become more populated when the aharmonicity,  $|\alpha|$, of the qubit becomes small. 

\begin{figure}[htbp]
\begin{center}
\includegraphics[width=8cm]{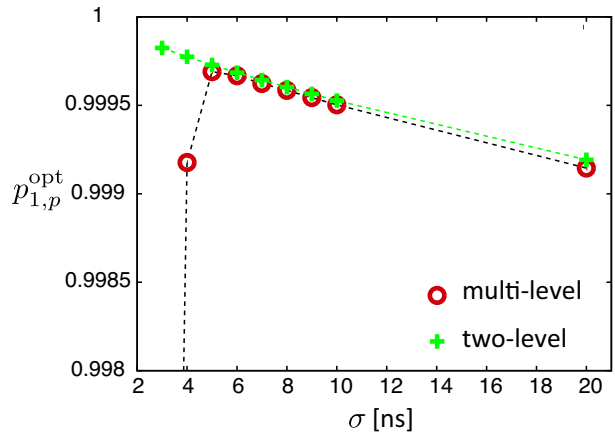}
\end{center}
\caption{
$p_{1,p}^{\rm opt}$ as a function of the pulse length $\sigma$.
The circles (crosses) are for the system with (without) higher levels.
The dashed lines are guide to the eye.
The other parameters used are the same as those in Fig.~\ref{pop_com_3_16_20}.
The four lowest levels of DQ and the two lowest levels of the JQF are taken into account.
(see Appendix \ref{Higer levels of JQF} for the results with more levels of the JQF.)
}
\label{max_fid_ver2_3_11_20}
\end{figure}

Figure~\ref{max_fid_ver2_3_11_20} shows $p_{1,p}^{\rm opt}$ as a function of $\sigma$.
The frequency and the amplitude of the drive field are optimized for each $\sigma$.
We have calculated $p_{1,p}^{\rm opt}$ also for the system where the DQ and the JQF are modeled as two-level systems ($|\alpha| \rightarrow \infty$), to highlight the decrease of $p_{1,p}^{\rm opt}$ in the case with higher levels.
For the system with two-level qubits,  $p_{1,p}^{\rm opt}$ decreases monotonically with respect to $\sigma$.
This is due to the following reason: the JQF does not protect the DQ from radiative decay while the control field is applied due to saturation of the JQF \cite{Koshino2020}.
Thus, the decay of the DQ is enhanced as the control pulse becomes longer.
In contrast, there is a peak of $p_{1,p}^{\rm opt}$ at $\sigma=5$~ns when the higher levels are taken into account.
When $\sigma<5$~ns, $p_{1,p}^{\rm opt}$ drops  because the
spectral width of the drive pulse becomes large and causes unwanted transitions to the higher levels of the DQ.
We have confirmed that $p_{1,p}^{\rm opt}$ is insensitive to the existence of the higher levels of the JQF
(see Appendix \ref{Higer levels of JQF}).
For $\sigma>5$~ns, the behavior of $p_{1,p}^{\rm opt}$ is similar to the case with two-level qubits, although it is slightly lower.

\subsection{comparison between cw and pulsed drives}
\begin{figure}[htbp]
\begin{center}
\includegraphics[width=7.5cm]{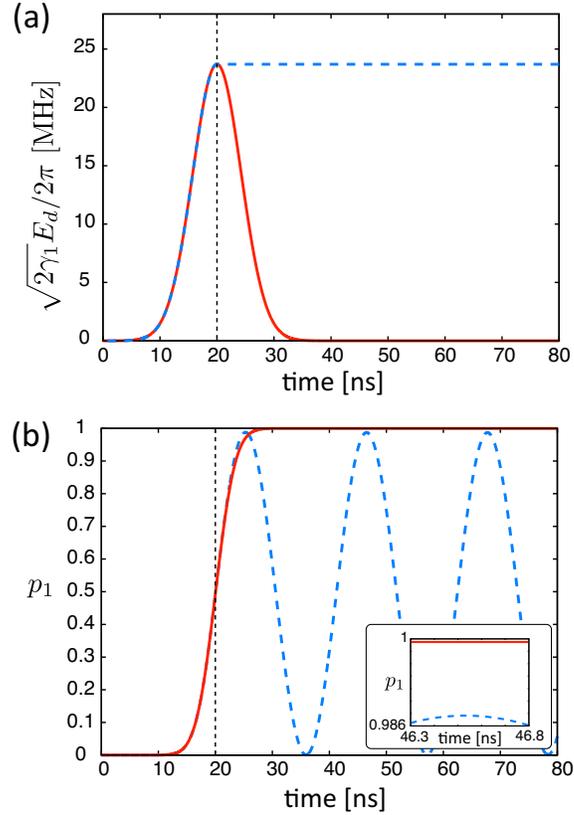}
\end{center}
\caption{
(a) Time dependence of $E_d$ in the controls with a cw field (blue dashed curve) and a gaussian pulse (red solid curve).
The vertical lines represent the position of the peak of the gaussian pulse.
(b) Time dependence of $p_1$ corresponding to $E_d$ in panel (a).
The inset is a closeup around the second peak of $p_1$ in the control with a cw field.
The used parameter set is $\sigma=10$~ns and
$\sqrt{2\gamma_1}E_{\rm amp}/2\pi=23.7$~MHz. We used the optimal drive frequencies of $5.0030$~GHz and $5.0024$~GHz for the controls with a cw field and a gaussian pulse, respectively.
The other parameters used are the same as those in Fig.~\ref{pop_com_3_16_20}.
}
\label{pop_Ed_4_5_20}
\end{figure}

We compare the controls with a cw field and a gaussian pulse of which $E_d$ is
defined in Eqs.~(\ref{Eamp_5_2_20}) and (\ref{Ed_gauss_5_4_20}), respectively.
Figures \ref{pop_Ed_4_5_20}(a) and \ref{pop_Ed_4_5_20}(b) show the time evolution of $E_d$ and $p_1$ during the controls.
The maximum $p_1$ for a gaussian pulse is higher than the one for a cw field.
In the control with a gaussian pulse, $p_1$ is sufficiently higher than 0.999, while it increases only up to 0.988 in the control with the cw field. This is because the narrow distribution of gaussian pulse in the frequency space decreases the effects of the higher levels of the DQ (The width of the pulse in the frequency space is narrower than the anharmonicity parameter),
while the population of $|2\rangle$ decreases $p_1$ in the control with the cw field as shown in Sec.~\ref{Effect of a higher level in cw drive}.

\section{Summary}
\label{Summary}
We have studied the effects of the higher levels of qubits on controls of the DQ protected by a JQF.
It has been shown that the higher levels of the DQ cause the shift of the resonance frequency and the decrease of the maximum population of the first excited state in the controls with a cw field and a pulsed field, while the higher levels of the JQF can be neglected.
The resonance frequency shift and the time evolution of the populations of the DQ under a cw field has been explained 
using a simplified model, which leads to simple formulae of the resonance frequency and the population matching well to the numerical results.
These results will be useful for the parameter determinations of the system with a cw field.

We have numerically examined the control with a pulsed field aiming at transferring the population to the first excited state
of the DQ from the ground state.
We have obtained the shift of the resonance frequency, which is inversely proportional to the anharmonicity parameter  similarly to the cw drive.
In contrast to the cw drive, the maximum population of the first excited state  
is insensitive to the anharmonicity parameter and is considerably higher the one of the cw drive, when the intensity of the anharmonicity parameter is sufficiently large.
The insensitivity of the control efficiency to the higher levels is attributed to the narrow distribution of the pulse field in the frequency space.
Moreover, we have shown optimal parameters of the pulsed field, which maximize the control efficiency.

\section*{Acknowledgments}
This work was supported in part by the Japan Society for the Promotion of Science (JSPS) Grants-in-Aid for Scientific Research (KAKENHI) (Grants No. 18K03486 and No. 19K03684), the Japan Science and Technology Agency (JST) Exploratory Research for Advanced Technology
(ERATO) (Grant No. JPMJER1601), and the Ministry of Education, Culture, Sports, Science, and Technology Quantum Leap Flagship Program (MEXT Q-LEAP) (Grant No. JPMXS0118068682).

\appendix

\section{Higer levels of JQF}
\label{Higer levels of JQF}
\begin{figure}
\begin{center}
\includegraphics[width=8cm]{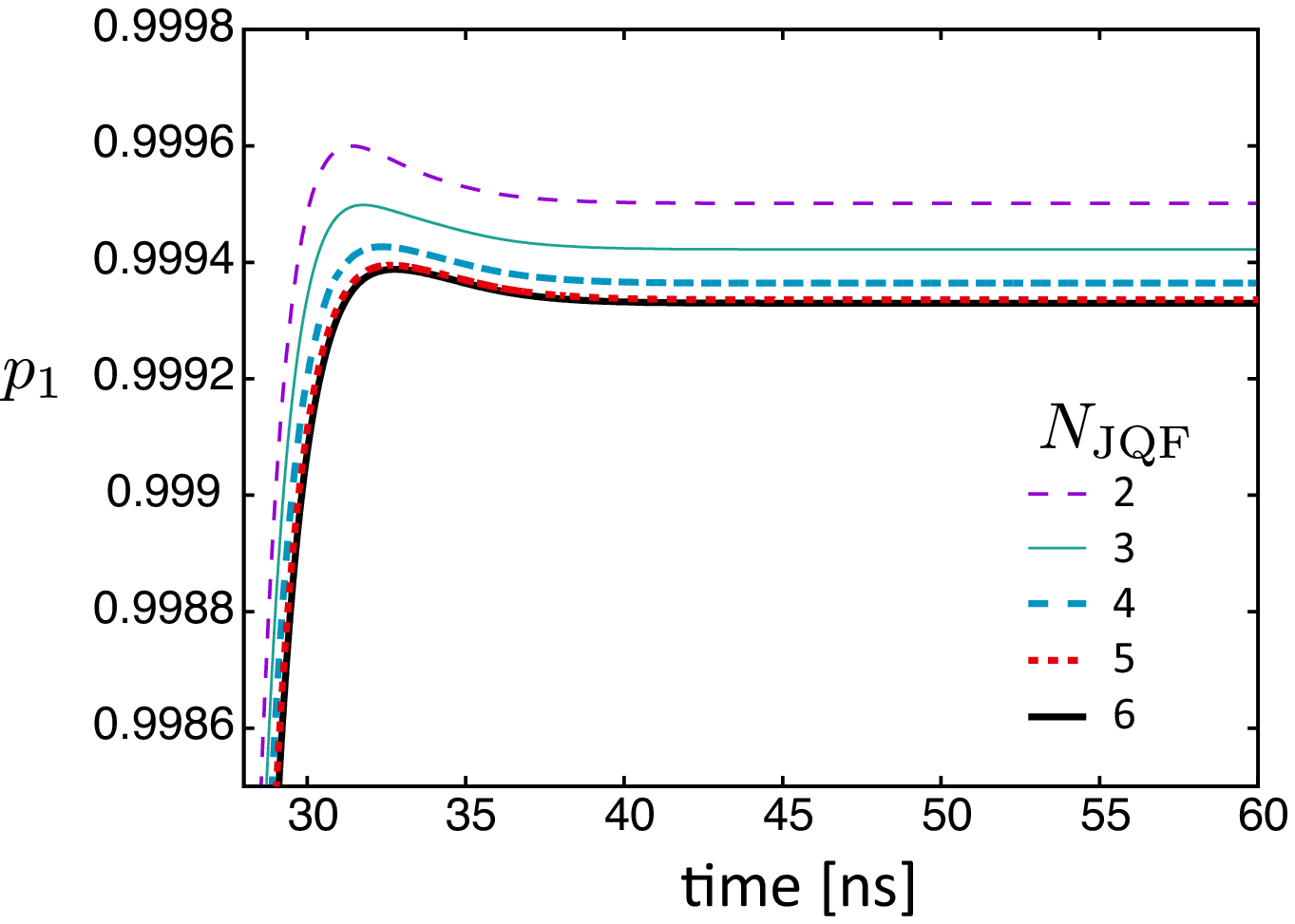}
\end{center}
\caption{
Time dependence of $p_1$ under the drive with a gaussian pulse, where
$N_{\rm JQF}$ is the number of the levels of the JQF taken into account.
The other parameters are the same as those in Fig.~\ref{dyn_gauss_3_17_20}.
}
\label{pop_com_7_13_20}
\end{figure}

We simulate the dynamics of the system under the drive with a gaussian pulse, taking into account the  higher levels of the JQF.
In our numerical simulations, we take into account $N_{\rm JQF}$ lowest levels of the JQF.
Figure~\ref{pop_com_7_13_20} shows the time dependence of $p_1$.
$p_1$ for larger $N_{\rm JQF}$ is slightly lower than the one for $N_{\rm JQF}=2$ due to the disturbance by the higher levels. 

\section{Analysis with Schrieffer-Wolff transformation}
\label{Analysis with Schrieffer-Wolff transformation}
We derive the shift of the resonance frequency and the decrease of maximum population of the first excited state in Rabi oscillations using the Schrieffer-Wolff transformation.

In the rotating frame at $\omega_{\rm d}$,
the Hamiltonian is represented as 
\begin{eqnarray}
H = H_0 + H_1 + V,
\end{eqnarray}
with
\begin{eqnarray}
H_0 &=& \varepsilon_1 \sigma_{11} + \varepsilon_2 \sigma_{22}\nonumber\\
H_1 &=&  \Omega (\sigma_{01} + \sigma_{10})\nonumber\\
V &=&  \sqrt{2} \Omega (\sigma_{12} + \sigma_{21}),
\end{eqnarray}
where we take into account only three levels $|0\rangle,|1\rangle$ and $|2\rangle$.
Here, $\varepsilon_1 = \omega - \omega_{\rm d}$, $\varepsilon_2 = 2(\omega-\omega_{\rm d})+\alpha$ and $\sigma_{ij}=|i\rangle\langle j|$. 
We transform the Hamiltonian (Schrieffer-Wolff transformation) as 
\begin{eqnarray}
H' &=& e^{-(S_1 + S_2)} H e^{S_1 + S_2}\nonumber\\
&=& H_0 + H_1 + V + [H_0,S_1] \nonumber\\
&&+ [H_1,S_1] + [V,S_1] + \frac{1}{2} [[H_0,S_1],S_1] + [H_0,S_2] + \cdots. 
\end{eqnarray}
We choose $S_{1,2}$ to diagonalize $H'$ up to $O(\Omega^2)$ except for $H_1$,
which is responsible for Rabi oscillation.
Here, $S_1$ is determined by $V+[H_0,S_1]=0$ as
\begin{eqnarray}
S_1 = -i \int^0_{-\infty} dt V(t) =
\frac{\sqrt{2}\Omega}{\varepsilon_2-\varepsilon_1} (\sigma_{12}-\sigma_{21}).
\end{eqnarray}
$S_2$ is determined by $[H_1,S_1] + [V,S_1] + \frac{1}{2}[[H_0,S_1],S_1] + [H_0,S_2]=0$,
which is rewritten as $[H_1,S_1]+[V,S_1]/2 + [H_0,S_2]=0$. 
Since $\frac{1}{2}[V,S_1] = \frac{2\Omega^2}{\varepsilon_2 - \varepsilon_1}(\sigma_{22}-\sigma_{11})$ is already diagonal,
we choose $S_2$ to satisfy $[H_1,S_1]+[H_0,S_2]=0$.
Since $[H_1,S_1] = \frac{2\Omega^2}{\varepsilon_2 - \varepsilon_1}(\sigma_{02}-\sigma_{20})$, we have
\begin{eqnarray}
S_2 &=& -i\int_{-\infty}^0 dt \frac{\sqrt{2}\Omega^2}{\varepsilon_2 - \varepsilon_1}
(\sigma_{02}e^{-i\varepsilon_2 t} + \sigma_{20}e^{i\varepsilon_2 t})\nonumber\\
&=& \frac{\sqrt{2} \Omega^2}{(\varepsilon_2 - \varepsilon_1) \varepsilon_2} (\sigma_{02} - \sigma_{20}).
\end{eqnarray}
Thus, $H' = H_0 + H_1 + [V,S_1]/2$ is given by
\begin{eqnarray}
H' &=& \Big{(} \varepsilon_1 - \frac{2\Omega^2}{\varepsilon_2- \varepsilon_1}\Big{)} \sigma_{11}
+\Big{(} \varepsilon_2 + \frac{2\Omega^2}{ \varepsilon_2 - \varepsilon_1}\Big{)} \sigma_{22} + \Omega (\sigma_{01} + \sigma_{10}).
\end{eqnarray}

Neglecting $\Omega(\sigma_{01} + \sigma_{10})$, the eigenstates of $H'$ are $|0\rangle$,
$|1\rangle$ and $|2\rangle$.
Thus, we have  
\begin{eqnarray}
H' |0\rangle &=& 0,\nonumber\\
H' |1\rangle &=& E_+ |1\rangle,\nonumber\\
H' |2\rangle &=& E_- |2\rangle,
\label{HS_7_21_20}
\end{eqnarray}
where $E_+ = \varepsilon_1 - 2\Omega^2/(\varepsilon_2 - \varepsilon_1)$ and $E_- = \varepsilon_1 - 2\Omega^2/(\varepsilon_2 - \varepsilon_1)$.
The resonance condition $E_+=0$ leads to the same form of the resonance frequency as Eq.~(\ref{Ep_7_21_20}).

Because Eq.~(\ref{HS_7_21_20}) is rewritten as 
\begin{eqnarray}
H e^{S_1 + S_2} |0\rangle &=& 0,\nonumber\\
H e^{S_1 + S_2} |1\rangle &=& E_+ e^{S_1 + S_2} |1\rangle, \nonumber\\
H e^{S_1 + S_2} |2\rangle &=& E_- e^{S_1 + S_2} |2\rangle,
\end{eqnarray}
the eigenstates of $H$ are 
\begin{eqnarray}
e^{S_1 + S_2}|0\rangle &=& |0\rangle,\nonumber\\
e^{S_1 + S_2}|1\rangle &=& |+\rangle,\nonumber\\
e^{S_1 + S_2}|2\rangle &=& |-\rangle.
\end{eqnarray}
Up to $O(\Omega)$, $|+\rangle$ is expanded as 
\begin{eqnarray}
|+\rangle = (1+S_1)|1\rangle = |1\rangle - \frac{\sqrt{2}\Omega}{\varepsilon_2 - \varepsilon_1} |2\rangle  + \cdots.
\end{eqnarray}
This means that the population of $|2\rangle$ in $|+\rangle$ is $(\frac{\sqrt{2}\Omega}{\varepsilon_2 - \varepsilon_1})^2$,
and the decrease of maximum population of the first excited state in Rabi oscillations is $(\frac{\sqrt{2}\Omega}{\varepsilon_2 - \varepsilon_1})^2$. This result is compatible with the one in Sec.~\ref{Effect of a higher level in cw drive}.

\end{document}